\documentclass[aps,pre,reprint,unsortedaddress,superscriptaddress]{revtex4-1} 

\usepackage{amsmath}  
\usepackage{amsfonts} 
\usepackage{graphicx} 

\usepackage{amssymb}

\usepackage{dcolumn}
\usepackage{bm}

\usepackage{verbatim}

\usepackage{color}
\usepackage{epstopdf}
\usepackage{hyperref}
\usepackage{xifthen}
\usepackage{changes}
\usepackage{ulem}

\DeclareMathAlphabet{\mathpzc}{OT1}{pzc}{m}{it}

\newcommand{\vect}{\boldsymbol}

\newcommand{\agg}{\text{a}} 
\newcommand{\mon}{\text{m}}

\newcommand{\map}{{M}_\agg^{\alpha}}

\newcommand{\cp}{{c}_\agg^{\alpha}}  
\newcommand{\cpI}{{c}_\agg^{\text{I}}}  
\newcommand{\cpII}{{c}_\agg^{\text{II}}}

\newcommand{\mam}{{M}_\mon^{\alpha}} 
\newcommand{\mamI}{{M}_\mon^{\text{I}}} 
\newcommand{\mamII}{{M}_\mon^{\text{II}}} 
\newcommand{\mamIeq}{{M}_{\mon,\text{eq}}^{\text{I}}} 
\newcommand{\mamIIeq}{{M}_{\mon,\text{eq}}^{\text{II}}} 

\newcommand{\mamt}{{M}_\mon^{\rm tot}}

\newcommand{\rb}{\beta}

\newlength\dlf  

\begin{document}



\title{
Spatial control of irreversible protein aggregation}

\author{Christoph A. Weber}
\affiliation{Engineering and Applied Sciences, Cambridge, Massachusetts 02138, United States of America}
\affiliation{contributed equally}

\author{Thomas C. T. Michaels}
\affiliation{Engineering and Applied Sciences, Cambridge, Massachusetts 02138, United States of America}
\affiliation{contributed equally}

\author{L. Mahadevan}
\affiliation{
Engineering and Applied Sciences, Physics and Organismic and Evolutionary Biology, Cambridge, Massachusetts 02138, United States of America}

\begin{abstract}
Liquid cellular compartments spatially segregate from the cytoplasm and can regulate aberrant protein aggregation, a process linked to several medical conditions, including Alzheimer's and Parkinson's diseases. Yet the mechanisms by which these droplet-like compartments affect protein aggregation remain unknown. Here, we combine kinetic theory of protein aggregation and liquid-liquid phase separation to study the spatial control of irreversible protein aggregation in the presence of liquid compartments. We find that, even for weak interactions between the compartment constituents and the aggregating monomers, aggregates are strongly enriched inside the liquid compartment relative to the surrounding cytoplasm. We show that this enrichment is caused by a positive feedback mechanism of aggregate nucleation and growth which is mediated by a flux maintaining the phase equilibrium between the compartment and the cytoplasm. Our model predicts that the compartment volume that maximizes aggregate enrichment in the compartment is determined by the reaction orders of aggregate nucleation. The underlying mechanism of aggregate enrichment could be used to confine cytotoxic protein aggregates inside droplet-like compartments suggesting potential new avenues against aberrant protein aggregation. Our findings could also represent a common mechanism for the spatial control of irreversible chemical reactions in general.
\end{abstract}



\maketitle 


\cleardoublepage

Spatial 
control within living cells is essential to many cellular activities, ranging from the local control of protein activity to the uptake of pathogens or the management of wastes~\cite{alberts2017molecular}. Understanding the mechanisms underlying regulation of cell activities in space and time is key not only for biological function, but also in view of understanding and eventually controlling cellular dysfunction~\cite{Knowles_2014,Chiti_2017,gitler2017neurodegenerative,michaels2018chemical}. 
The spatial organization of cellular activities is often associated with membrane-bound organelles that ensure permeation only for certain molecules of specific molecular structure~\cite{neupert2007translocation, wiedemann2017mitochondrial, 	dukanovic2011multiple}.
Recently, new types of organelles have been discovered that do not possess a membrane. They are referred to as non-membrane bound compartments and they share most hallmark properties with actual liquid-like droplets~\cite{brangwynne2009germline, brangwynne2013phase, elbaum2015disordered, zhu2015nuclear, banani2017biomolecular}. 
Unlike organelles surrounded by membranes, these non-membrane bound compartments are formed by liquid-liquid phase separation.  In many cases this phase separation is driven by disfavouring interactions between the constituent molecules of the compartment and the surrounding cytoplasm~\cite{hyman2014liquid,brangwynne2015polymer}. 
The partitioning of other intracellular molecules into such droplet-like compartments is then controlled by their relative interactions with the constituent molecules of the compartment. 

 \begin{figure}[tb]
	\begin{center}
		\includegraphics[width=0.5\textwidth]{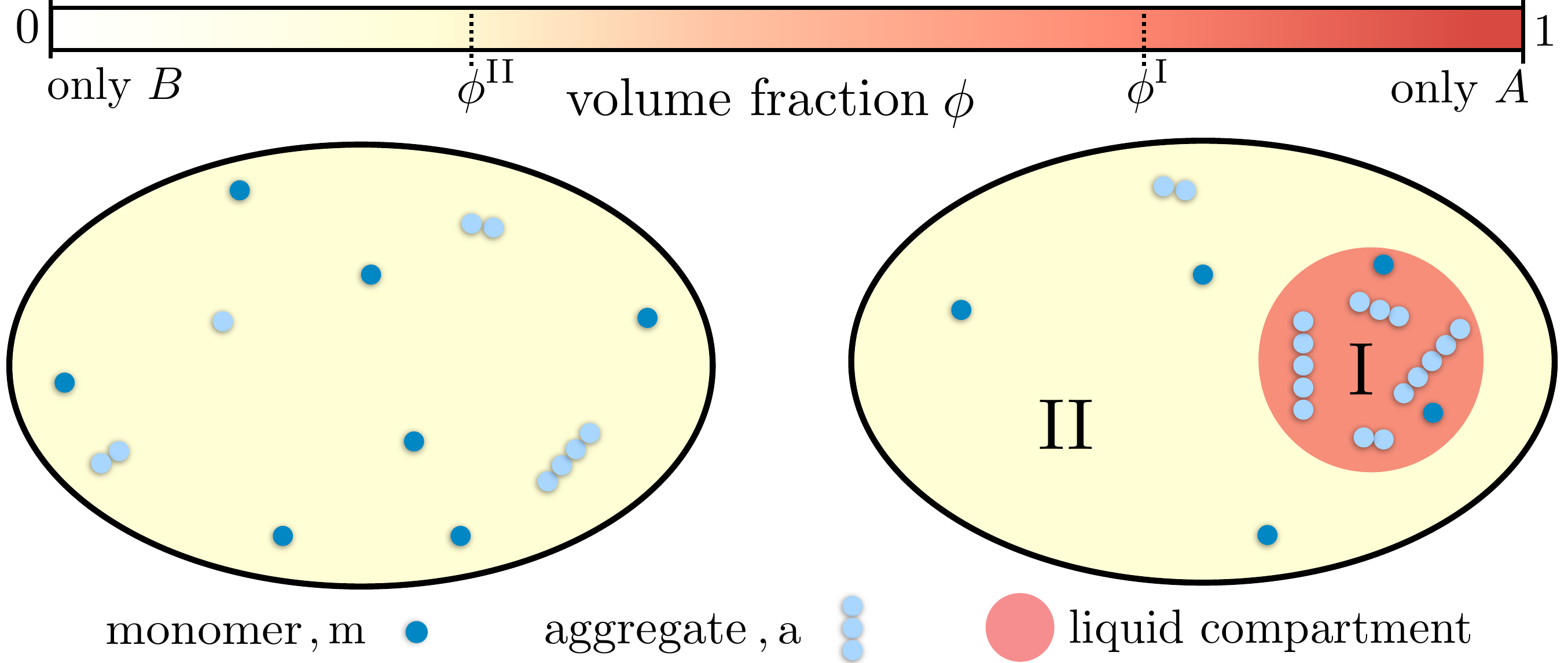}
	\end{center}
	\caption{
{\textbf{Enrichment of monomers and aggregates via liquid-like compartments.}} 
Protein aggregation may occur homogeneously inside cells also leading to aggregates inside more sensitive cellular regions (left). 
A liquid compartment may accumulate monomers and thereby trigger the local formation of aggregates (right). The hardly diffusing aggregates are thus kept away from a more sensible cellular region.
Such a spatial segregation of aggregates is ideal for adding functional, drug-like molecules 
which dominantly dissolve inside the compartment.
These molecules may degrade the aggregates or inhibit further growth and nucleation.
But most importantly, as these molecules are localized inside the compartment their toxic effects are diminished.
}
	\label{fig_1_illustration}
\end{figure}

 These droplet-like compartments are ubiquitous inside living cells~\cite{banani2017biomolecular}.
 For instance, they emerge prior to cell division~\cite{brangwynne2009germline, parker2007p}, 
 and form as a response to cellular  stress~\cite{Patel_liquid_to_solid_2015,Malinovska_Kroschwald_Alberti_2013,molliex2015phase}.
 They have been shown to enrich proteins~\cite{hernandez2017local,woodruff2017centrosome,mateju2017aberrant} and genetic material~\cite{parker2007p, saha2016polar, zhang2015rna} 
 providing distinct environments for chemical reactions and biological function. 
  The molecules hosted inside these compartments may even be protected against other agents from the cytoplasm~\cite{franzmann2018phase} or face conditions facilitating their molecular repair~\cite{mateju2017aberrant, ganassi2016surveillance, alberti2017granulostasis,
alberti2018quality,  jain2016atpase, specht2011hsp42}.
In addition to these roles, recent evidence suggests that liquid cellular compartments could play an important role in regulating pathological protein aggregation~\cite{alberti2016aberrant,shin2017liquid}.  An example is the irreversible assembly of amyloids into fibrillar aggregates, a process that is linked to a large variety of currently incurable diseases~\cite{dobson2003protein,Knowles_2014,lashuel2002neurodegenerative,catalano2006role,benilova2012toxic,campioni2010causative}, such as Alzheimer's and Parkinson's diseases, amyloidosis or type-II diabetes. As another example,  a chaperone in yeast uses a prion-like, intrinsically disordered domain to bind and sequester misfolded proteins in protein deposition sites~\cite{grousl2018prion,boczek2018one}. Moreover, misfolded and pathological proteins can  accumulate inside liquid-like stress granules triggering the aggregation kinetics inside these compartments.   The presence of this phase separated compartment  can promote the formation fibrillar  aggregates, and prevent aggregation outside the stress granules~\cite{molliex2015phase,mateju2017aberrant}. Thus the corresponding cytotoxic effects of protein aggregates are expected to be strongly localized in space  as well. However, it remains elusive whether weak protein interactions are sufficient to cause a significant change in aggregate concentration in the compartment relative to homogeneous aggregation and how the physical parameters of aggregation and phase separation determine the 
relative enrichment of aggregates inside versus outside.

Here, we combine knowledge of the kinetics of irreversible protein aggregation with the theory of liquid-liquid phase separation to develop a model of irreversible assembly of protein fibrils  in the presence of droplet-like compartments.  We use this model to predict the degree of enrichment of aggregates into the liquid compartment as a function of the fundamental physical parameters underlying aggregation kinetics and phase separation. We find that relatively weak interactions between the protein monomers and the liquid compartment molecules are sufficient to enrich the concentration of aggregates within the liquid compartment by several orders of magnitudes relative to homogeneous aggregation (Fig.~\ref{fig_1_illustration}). This strong enrichment of aggregates emerges because the liquid compartment acts as continuous sink of monomers during the aggregation dynamics, thus promoting intra-compartment aggregation but suppressing aggregation outside of the compartment. Moreover, we find that aggregate enrichment is more pronounced for larger (smaller) compartments depending on the relative values of the reaction orders for primary and secondary nucleation. Our results suggest that cellular liquid compartments are ideal to control irreversible protein aggregation in space and that the compartment volume, which is determined by the mean concentration of phase separated protein, represents a relevant control parameter for intra-compartment positioning of aggregates. This control mechanism could be relevant in the context of spatial regulation of other irreversible chemical reactions where liquid compartments act as biomolecular microreactors.

\section*{Mathematical model for liquid compartments controlling protein aggregation}
 
To capture the interplay between liquid phase separation and protein aggregation kinetics we start with a model of two coexisting protein phases  where monomers partition differently into  these phases determined by their relative interactions. 
 We consider the case where the partitioning of monomers is close to equilibrium continuously during the kinetics of aggregation. This assumption is well justified since small weakly interacting molecules such as the aggregating monomers diffuse between seconds and minutes through a  cell of size in the order of tens of $\mu m$~\cite{brangwynne2009germline,griffin2011regulation}, while typical time scales of aggregation \emph{in vitro} are in the order of hours (see e.g.~\cite{cohen2013proliferation}). Furthermore, the diffusion of aggregates is highly hindered as long fibrillar aggregates experience a much larger hydrodynamic drag force and can get entangled with cytoskeletal filaments and other assembled fibrils~\cite{de1971reptation,rubinstein1987discretized}. Finally, at large enough density and size, fibrils start to form solid-like gels~\cite{mateju2017aberrant} further slowing down their mobility. All these effects imply that we may safely neglect diffusion of large aggregates and consider the typical case that monomers diffuse quickly relative to their aggregation kinetics. This separation of time scales suggests that it is natural to first consider the partitioning of monomers into phase separated compartments at equilibrium and then consider small deviations from this equilibrium to understand its consequences for protein aggregation. 
 
\subsection*{Phase separation and partitioning of monomers at equilibrium}

 \begin{figure}[t]
\centering
\includegraphics[width=0.5\textwidth]{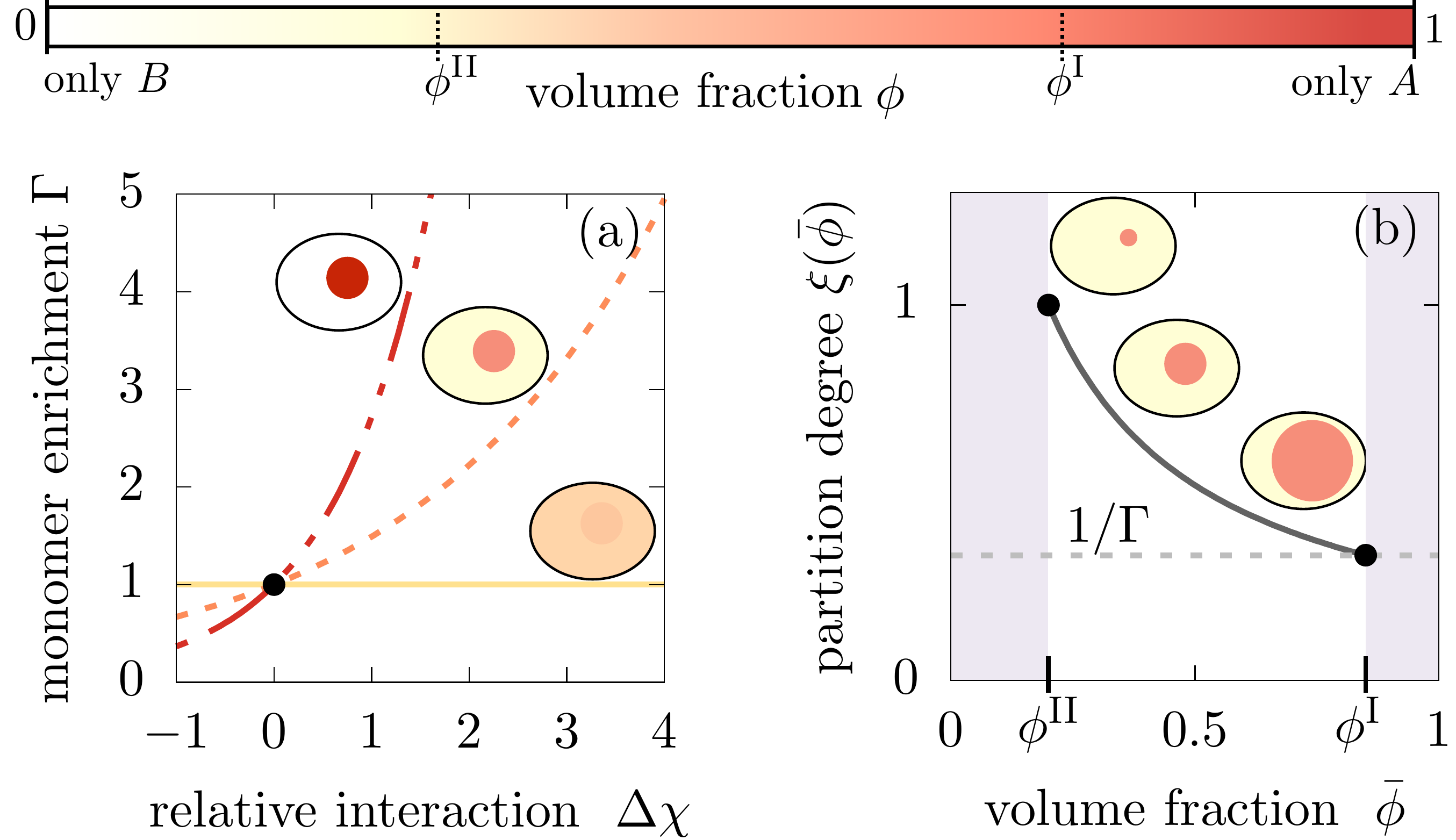} 
	\caption{
{\textbf{Enrichment of monomers and relative degree of segregation.}}
(a) The monomer enrichment $\Gamma$ (\eqref{relation_Gamma})
exponentially increases with the relative interaction strength $\Delta \chi$ (units of $k_BT$) between the monomers and the $A$ and $B$ molecules which is defined in the SI (section~S1). 
Its characteristic increase it set by the degree of phase separation, $\phi^\textrm{I} - \phi^\textrm{II}$. 
Enrichment vanishes at the critical point of phase separation (solid line)
and increases with the degree of phase separation (dashed line).
Enrichment is largest for $\phi^\textrm{I} - \phi^\textrm{II}\simeq1$ (dash-dotted line).
Due to the exponential increase, large enrichment of monomers $\Gamma$ can already be reached for weak relative interaction energies of a few $k_\textrm{B}T$.
(b) The partition degree $\xi(\bar \phi)= c_{\rm m}^\textrm{II} / c_\textrm{m}^\textrm{tot}$  (\eqref{relation_xi}) describing the concentration fraction of monomers that resides in the minority phase II of the compartment,  decreases with the mean volume fraction of $A$ material, $\bar \phi$, along with increasing compartment volume $V^\textrm{I}(\bar \phi)$.
Smaller compartments are thus better in enriching the monomer mass concentration. 
}
	\label{fig_2_monomer_volume_enrichment}
\end{figure}

We consider a system of total volume $V$ hosting a single liquid compartment (a droplet for example) of a condensed protein phase I of volume $V^\textrm{I}$. The compartment itself forms by liquid-liquid phase separation between the two components $A$ and $B$. Compartment I is composed mostly of the protein component $A$ and a small fraction of protein component $B$, while compartment II   has a small amount of protein $A$  and a large amount of protein $B$, as depicted in Fig.~\ref{fig_1_illustration}.

For simplicity, we discuss the case of an incompressible system where the aggregating monomers `m' and aggregates `a' are dilute, i.e., $c_{\rm m}\nu_\textrm{m} \ll1$ and $c_{\rm a}\nu_\textrm{a} \ll1$, with $c_\textrm{m}$ and $c_\textrm{a}$ denoting the concentrations of monomers and aggregates and $\nu_\textrm{m}$ and $\nu_\textrm{a}$ are the respective molecular volumes. For instance, typical values of volume fractions for monomers of Amyloid-$\beta$, $c_{\rm m} \nu_{\rm m}$ (radius of gyration in the range $1$--2 nm~\cite{sajfutdinow2018direct}), at physiological concentrations between 100pM to 1nM are in the range of $10^{-9}$ to $10^{-8}$. The assumption of dilute monomers and aggregates implies that for an incompressible system  the volume fractions $\phi_A$ and $\phi_B$ of the protein components $A$ and $B$ obey $\phi_A+\phi_B = 1 - c_{\rm m} \nu_{\rm m} - c_\textrm{a} \nu_\textrm{a}  \simeq 1$, where we abbreviate $\phi_A=\phi$ in the following. The monomers may partition differently into the respective minority and majority phases, but, due to their dilute concentrations, they do not affect the degree of phase separation. Under these circumstances, the partitioning of monomers in the two phases is solely governed by the relative interaction  strength $\Delta \chi$
between the monomers with the $A$ and the $B$ components, respectively. 
If $\Delta \chi$ is large and positive, monomers favor the presence of the majority component $A$ in compartment I.  
In this case we expect a more pronounced partitioning of monomers into compartment I. Contrariwise, when $\Delta \chi$ is large and negative, monomers favorably partition into compartment II. The degree of monomer enrichment  at equilibrium can be calculated using the condition that the chemical potentials of monomers associated with compartment I and II are balanced (see Fig.~\ref{fig_2_monomer_volume_enrichment}(a), SI, section~S1, for the derivation), and allows us to define the monomer enrichment factor
 \begin{equation}\label{relation_Gamma}
\Gamma \equiv \frac{c_{\rm m}^\textrm{I}}{c_{\rm m}^\textrm{II}}
\simeq  \exp\left[\frac{\nu_{\rm m}}{\nu}\Delta \chi  \left( \phi^\textrm{I} - \phi^\textrm{II}\right) \right]   
 \, ,
\end{equation}
where $c_{\rm m}^\textrm{I}$, $c_{\rm m}^\textrm{I}$ are the monomer concentrations in phases I and II, respectively, $\nu$ denotes the molecular volume of $A$ and $B$ molecules, and $\phi^\textrm{I} - \phi^\textrm{II} \in [0,1]$ is the degree of phase separation of the $A$-component. Then the relative  partitioning of the total monomer concentration, $c_\textrm{m}^\textrm{tot}= \left( c_{\rm m}^\textrm{I} V^\textrm{I} + c_{\rm m}^\textrm{II} V^\textrm{II} \right) /V$,  is given by the expressions  
$c_{\rm m}^\textrm{I} =\xi(\bar{\phi}) \Gamma c_\textrm{m}^\textrm{tot}$ and $c_{\rm m}^\textrm{II} =\xi(\bar{\phi})  c_\textrm{m}^\textrm{tot}$,
where the partition degree
\begin{equation}\label{relation_xi}
 	\xi\left( \bar \phi \right)=\frac{1}{1+ (\Gamma -1)V^\textrm{I}\left(\bar \phi \right)/V} 
\end{equation}
 captures the impact of the relative size of the compartment volume $V^\textrm{I}( \bar{\phi} )/V$. The volume of the compartment I is in turn controlled by the mean volume fraction $\bar \phi$ of $A$ molecules in the system in terms of the relationship $V^\textrm{I}(\bar \phi)= V (\bar \phi - \phi^\textrm{II})/( \phi^\textrm{I} - \phi^\textrm{II})$.
The relationships~\eqref{relation_Gamma} and \eqref{relation_xi} characterize the relatively rapid physics of monomer partitioning at equilibrium and serve as the background atop which we consider out-of-equilibrium aggregation kinetics.

\subsection*{Model for protein aggregation kinetics coupled to non-equilibrium monomer partitioning}

 Due to the separation of time scales of monomer diffusion and monomer aggregation, the partitioning of monomers into the compartment is close to  equilibrium at all times of the aggregation kinetics and thus the relative fraction of monomers is approximately governed by monomer enrichment $\Gamma$, \eqref{relation_Gamma}. However, as the aggregation kinetics decreases the amount of monomers inside each phase, aggregation  couples to the partitioning which cause the generation of diffusive fluxes  $J^\alpha$  that attempt to maintain the monomer partitioning close to equilibrium. In the limit of a sharp interface separating the liquid compartment from the bulk, so that there is no aggregation at the interface,  $J^\textrm{I}=-J^\textrm{II}$. 
Furthermore, to linear order, the flux ${J^\alpha}$  between the phases is  proportional to the difference of monomer partitioning with respect the equilibrium value $\Gamma$ (see SI, section~S2, for the derivation) and is of the form: 
 \begin{subequations}\label{1133}
   \begin{align}\label{eq_flux}
 	\frac{J^\alpha}{V^{\alpha}(R)} & = - k^{\alpha} \,  \left(  \mamI - \Gamma  \mamII \right) \, ,
 \end{align} 
  where  
  $\mam=c_{\rm{m}}^\alpha m_\textrm{m}$ (with $m_\textrm{m}$ as monomer mass) is the monomer mass concentration
  in compartment  $\alpha=\textrm{I},\textrm{II}$, and 
  $k^\alpha$ 
 denotes the rate at which monomer partitioning relaxes back to the equilibrium given by \eqref{relation_Gamma}. 
  
In each phase,  irreversible protein aggregation of fibrillar structures occurs as a consequence of both primary and secondary nucleation, and growth of aggregates via their ends, each event occurring with a rate $k_1$, $k_2$, and $k_+$~\cite{michaels2014mean,michaels2016hamiltonian,arosio2016kinetic,michaels2018chemical}. We see that the key term in our model is the difference between the monomer concentration inside and outside of the compartment which leads to the diffusive flux of monomers $J^\alpha$ between the phases (\eqref{eq_flux}), which connects the effects of phase separation and protein aggregation. The coupled equations describing the protein aggregation in both phases can
 be written as
\begin{align}
\label{1133c}
\frac{\textrm{d}\cp(t)}{\textrm{d}t} & = k_1 \, \mam(t)^{n_1}+k_2 \, \mam(t)^{n_2} \, \map(t) \, ,\\
\label{1133b}
 \frac{\textrm{d}\map(t)}{\textrm{d}t} & =  2k_+\, \mam(t) \, \cp(t)  \, ,\\
 \label{1133a}
\frac{\textrm{d}\mam(t)}{\textrm{d}t} &= -2k_+\, \mam(t) \, \cp(t)  + \frac{J^{\alpha}}{V^\alpha} \, .
 \end{align} 
 \end{subequations}
%
%
%
Here, \eqref{1133c} describes the formation of new fibrils in each compartment through primary nucleation, fragmentation or surface catalyzed secondary nucleation,  \eqref{1133b} captures the buildup of aggregate mass in each compartment due to elongation of existing aggregation by monomer addition, and \eqref{1133a} models the population balance of monomers in each compartment as a result of aggregate growth and the flux monomer between compartments I and II. This flux is given by (\eqref{eq_flux}) ensuring that partitioning is maintained close to the monomer enrichment factor $\Gamma$.

While the monomer enrichment factor $\Gamma$ (\eqref{relation_Gamma}) governs the constant ratio of the time dependent concentrations in compartment I and II, the partitioning degree $\xi$ (\eqref{relation_xi}) determines how the total monomer concentration, which decays over time as a result of aggregation, is split between the two compartments at any time point during the kinetics of aggregation. 
As we will see, both parameters will be crucial in controlling the 
degree of aggregate enrichment inside the compartments. 

\section*{Irreversible aggregation in the presence of phase separated compartments}

To understand how protein aggregation kinetics couples to two phase separated compartments in terms of the physical parameters $\Gamma$ and $\xi$, we constructed explicit analytical solutions to the set of non-linear kinetic equations~\eqref{1133} by exploiting an analogy to classical mechanics~(\cite{michaels2016hamiltonian} and  SI, section~S3, for details), and compared with numerical solutions of \eqref{1133}.

\subsection*{Monomer enrichment causes a relative change in nucleation and growth of aggregates between the compartments}

In the limit of fast monomer diffusion the aggregation kinetics in each compartment is controlled by a set of effective rate parameters.
 The  relative magnitude of these effective rates between compartment I and II at early times 
 scales with the monomer enrichment as $\Gamma^{n_1}$,
while at late times, the corresponding ratio of these rates scales with $ \Gamma^{(n_2+1)/2}$
 (see SI, Eq.~(S33) and~(S34)). 
Thus, the aggregate growth inside compartment I is faster than in compartment II  if there is enrichment of monomers in the condensed phase, i.e., when $\Gamma>1$. Moreover, the relative magnitudes of growth rate at early times solely depends on the reaction order of primary nucleation, $n_1$,  while at late times, relative growth is determined by the reaction order of secondary nucleation, $n_2$.

 \begin{figure*}[tb]
\centering
\includegraphics[width=1.0\textwidth]{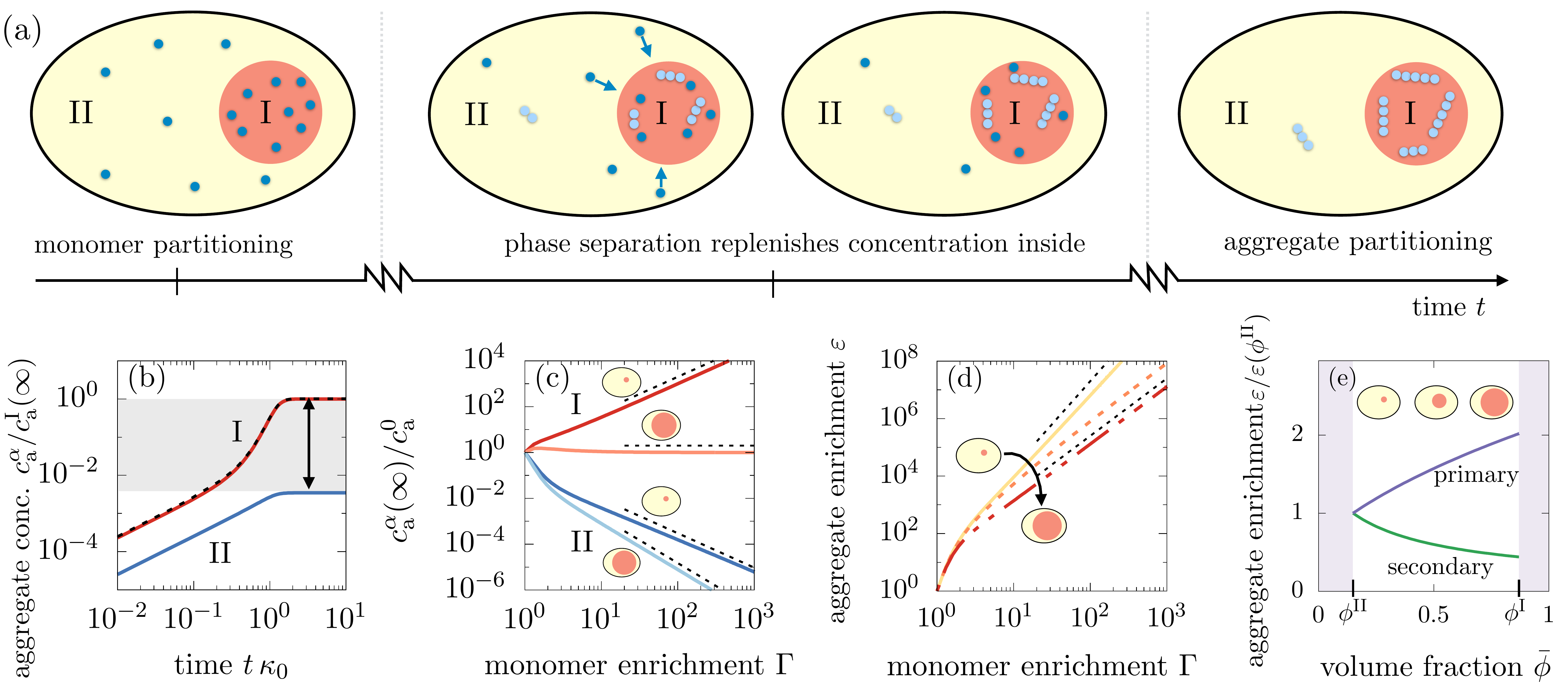} 
	\caption{
	\textbf{Segregation of aggregates into compartment I via positive feedback mediated by phase separation.} 
	(a) Sketch of aggregation kinetics inside the two compartments I and II. 
	Left: Initially, monomers get enriched on a short diffusive time scales due to the partitioning mediated by the phase separated compartments (\eqref{relation_Gamma}).  
	Center: Monomers slowly aggregate. More aggregates nucleate 
	and grow in compartment I due to the initial partitioning of monomers. This pronounced, initial aggregation causes a continuous monomer flux into compartment I, further promoting aggregation (positive feedback indicated by arrows). 
	Right:  Partitioning of monomers together with the positive feedback can cause a very pronounced accumulation of aggregates relative to compartment II. 
	(b) Aggregate concentration $\cp(t)$ as a function of time $t$ obtained from solving numerically and analytically Eqs.~\eqref{1133} actually confirms that  aggregates can enrich by several orders of magnitude. 
	(c) The asymptotic concentrations $c_{\rm a}^\textrm{I}(\infty)$ and $c_{\rm a}^\textrm{II}(\infty)$ inside each of the compartment inversely scale  for small compartments, while for large compartment I, aggregate enrichment therein vanishes while depletion inside compartment II is dominated by primary nucleation.The asymptotic concentration in the absence of monomer enrichment, $\Gamma=1$, is denoted as $c_\textrm{a}^0$. Dashed line are the scalings given in the the main text. Parameters: $n_1=n_2=2$.
	 (d) Enrichment factor $\varepsilon$ of aggregates inside compartment I as a function of monomer enrichment $\Gamma$ can reach very large values. The behaviour switches from secondary nucleation dominated increase at small compartment I volumes to primary  dominated growth at large volumes. 
	 Dashed line are the scalings given in \eqref{enrichment}. 
	 (e) The slope of the enrichment factor as a function of mean volume fraction $\bar \phi$, equivalently speaking, volume of compartment I, changes its sign when enrichment is dominated by primary ($n_1=2, n_2=0$) or secondary nucleation ($n_1=2, n_2=2$). 
	Parameters: (b,e) $\Gamma=3$ consistent with weak interactions. 
}
	\label{fig_3_monomer_volume_enrichment}
\end{figure*}

\subsection*{Phase separated compartments mediate a positive feedback for aggregate growth}

This difference in growth rates between the phases can be qualitatively explained by the rapid  preference of monomers to recover phase equilibrium  (Fig.~\ref{fig_3_monomer_volume_enrichment}(a)). The higher monomer concentration in compartment I causes  aggregates to nucleate first inside compartment I. As a consequence elongation of aggregates is more pronounced inside compartment I leading to a stronger consumption of monomers. This difference in monomer consumption between the compartments couples to the flux \eqref{1133}, which forces more monomers to diffuse into compartment I to maintain partitioning equilibrium, even as aggregates grow.  This positive feedback mechanism in compartment I is accompanied by negative feedback for compartment II, which continuously looses monomers leading to a slow down of the aggregation kinetics outside. Thus, the coupling between the aggregation kinetics and phase separation, mediated by \eqref{1133}, is key to determine aggregate enrichment/depletion in each phase.
 
 \subsection*{Positive feedback for aggregate growth causes strong  aggregate enrichment}

To understand this feedback mechanism we study the time evolution of the aggregate concentration inside each phase, $\cpI(t)$ and $\cpII(t)$ (Fig.~\ref{fig_3_monomer_volume_enrichment}(b)). As aggregation is initiated by primary nucleation, it is solely determined by the monomer concentration. Because monomer concentrations in the compartments are slaved due to the rapid flux that maintains partitioning equilibrium, the time evolution of the aggregate concentrations  in the early regime of the aggregation kinetics
are slaved as well, following $\cpI(t)/\cpII(t) \propto \Gamma^{n_1}$. When aggregates start consuming monomers via elongation, 
the flux of monomers from compartment II to I causes a saturation of the  aggregate concentration outside the compartment II, while the concentration of aggregates in compartment I increases significantly. This rapid increase of growth is facilitated by the continuous influx of monomers (positive feedback). As monomers get depleted in the entire system the growth of aggregates also saturates in compartment I. Most importantly, the resulting asymptotic concentrations at large time scales, 
$c_{\rm a}^\textrm{I}(\infty)$ and $ c_{\rm a}^\textrm{II}(\infty)$, can differ by several orders of magnitude, even for modest values of $\Gamma$ corresponding to weak relative interactions.

 \subsection*{Enrichment and depletion relative to homogeneous aggregation is determined by the reaction orders}
 
To elucidate the impact of the reactions orders on the aggregation kinetics
we first consider the enrichment of aggregates relative to the case of homogeneous  aggregation, i.e., for $\Gamma=1$. For large values of monomer enrichment, the asymptotic concentrations in compartments I and II at large times relative to the homogeneous aggregate concentration $c_\textrm{a}^0$ at large times (see SI, Eq.~(S46), for the definition) read
\begin{align}
\label{cII_asymptotic}
 \frac{c_{\rm a}^\textrm{II}(\infty)}{c_\textrm{a}^0 w} &\simeq  \xi(\bar{\phi})^{{n_1}-\frac{n_2+1}{2}} \, \Gamma^{-\frac{n_2+1}{2}} \, ,\\
\label{cI_asymptotic}
\frac{c_{\rm a}^\textrm{I}(\infty)}{c_\textrm{a}^0} &\simeq \left(\xi(\bar{\phi})\,\Gamma \right)^{\frac{n_2+1}{2}} \, ,
\end{align}
where $w$ is a dimensionless numerical prefactor (see SI, Eq.~(S49)).
%
We see that for a large monomer enrichment factor $\Gamma$, 
the enrichment of aggregates inside compartment I gets more pronounced, while aggregates in compartment II are more depleted relative to the homogeneous case (Fig.~\ref{fig_3_monomer_volume_enrichment}(c)). 
Most impotantly, the value of the aggregate concentration for given monomer enrichment is controlled by the reaction orders for primary and secondary nucleation, $n_1$ and $n_2$.

 \subsection*{Aggregate concentration in the compartments is controlled by compartment volume}

Having understood the role of the monomer enrichment factor $\Gamma$ in aggregation kinetics, we now turn to 
how the asymptotic concentrations of aggregates in each compartment depend on the volumes of the compartments. The dependence on compartment volume is given the partition degree $\xi(\bar{\phi})$. From  \eqref{relation_xi}, we see that the partition degree $\xi(\bar{\phi}) \in [1, \Gamma^{-1}]$, where the value of one is relevant for small compartments (Fig.~\ref{fig_2_monomer_volume_enrichment}(b)). Following \eqref{cII_asymptotic} and \eqref{cI_asymptotic}, we see that for a small volume of compartment I,  enrichment and depletion exhibit an inverse scaling with 
$c_{\rm a}^\textrm{I}(\infty) \propto \left( c_{\rm a}^\textrm{II}(\infty)\right)^{-1} \propto \Gamma^{\frac{n_2+1}{2}}$, which is solely dependent on the reaction order for secondary nucleation. Contrariwise, when the volume of compartment I is large,  enrichment of aggregates inside I vanishes, while depletion inside compartment II then solely depends on the reaction order for primary nucleation, $c_{\rm a}^\textrm{II}(\infty) \propto \Gamma^{-n_1}$.

This switch between aggregate enrichment governed by secondary nucleation, to an enrichment solely determined by primary nucleation,
arises from
primary nucleation events occurring first inside compartment I
 due to a higher monomer concentration ($\Gamma>1$). 
 Once the first aggregates have formed via primary nucleation inside compartment I, small and large compartments behave fundamentally differently. If compartment I is small, only a few aggregates can form via primary nucleation due to the small compartment size. As aggregates begin to grow earlier in compartment I, the unbalance of monomers causes a flux from II to I. As a consequence of this continuous flux, the   secondary nucleation events quickly overwhelm  primary nucleation events inside compartment I, while  
 secondary nucleation is suppressed in compartment II. 
 However, if compartment I is large, the aggregation kinetics is similar 
 to that for a homogeneous system because the monomer mass concentration is very close to the total monomer mass in the system and there is only a negligible amount of monomers entering from compartment II.  Additionally, in the smaller compartment II where aggregates grow via primary nucleation,  the coupling flux  continuously removes monomers suppressing primary nucleation. Since compartment I is large, it shows little or no enrichment of aggregates relative to the homogeneous case while inside the small compartment II, aggregates are depleted determined by the lack of primary nucleation 
 events relative to the homogeneous case.

\subsection*{Changes in compartment volume switch the driving  mechanism for aggregate enrichment}

 \begin{figure}[tb]
\centering
\includegraphics[width=0.42\textwidth]{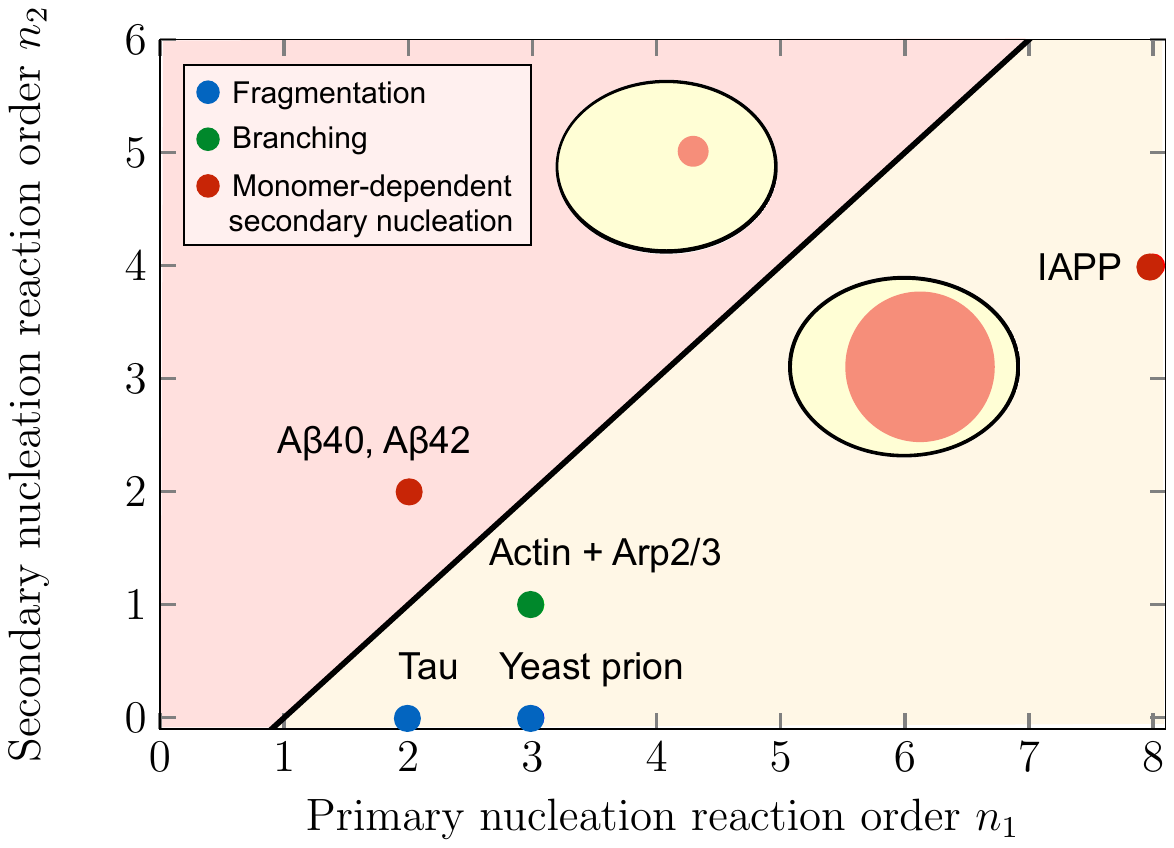} 
	\caption{
	\textbf{Theoretical predictions of maximal aggregate enrichment for various aggregating systems.}
	Our predictions are summarized by a phase diagram depicting that aggregating systems characterized by different reaction orders for primary and secondary nucleation, $n_1$ and $n_2$, 
 show maximal aggregate enrichment for large or small compartments, respectively.
	 The two regions where either large or small compartments lead to a larger enrichment of aggregates is separated by the line $n_2 = n_1-1$ determined from \eqref{enrichment}. 
	 For $n_2 > n_1-1$ small compartments lead to larger aggregate enrichment, while for $n_2 < n_1-1$, larger compartments are beneficial. To illustrate which scenario might apply to which kind of aggregating system, we indicate the measured values of the primary and secondary reaction orders for a range of systems propagating through fragmentation (blue), lateral branching (green) or monomer-dependent secondary nucleation (red): 
	 Tau~\cite{kundel2018measurement}, 
	 yeast prion Ure2p~\cite{zhu2003relationship}, IAPP~\cite{ruschak2007fiber}, amyloid-$\beta 40$ (for monomer concentrations below 5 $\mu$M)~\cite{meisl2014differences}, 
	 amyloid-$\beta 42$~\cite{cohen2013proliferation}. 
}
\label{fig_4_phase_diagram_large_small_compartments}
\end{figure}

To quantify the switch in aggregation enrichment as a function of compartment volume we define the asymptotic aggregate enrichment ratio
\begin{equation}\label{enrichment}
\varepsilon(\bar{\phi})=\frac{c_{\rm a}^\textrm{I}(\infty)}{c_{\rm a}^\textrm{II}(\infty)}
   \propto   \xi(\bar{\phi})^{n_2-n_1+1} \, \Gamma^{n_2+1} \, .
\end{equation}
 As the compartment volume enters the enrichment factor $\varepsilon(\bar{\phi})$ solely via the partition degree $\xi(\bar{\phi})$, the sign of $n_2-n_1+1$ determines whether larger or smaller compartments lead to a larger enrichment (Fig.~\ref{fig_3_monomer_volume_enrichment}(e)). Indeed we find that the slope of the enrichment factor scales as $\varepsilon(\bar{\phi})^\prime \propto (n_1-n_2-1)$. 
Thus, for $n_1 > n_2+1$, increasing the compartment volume by increasing the amount of $A$-material $\bar{\phi}$  causes a larger relative enrichment. Conversely, for $n_1 < n_2+1$, larger enrichment can be found for smaller compartment sizes.
Consistently, if the nucleation coefficients obey $n_2 = n_1-1$, compartment volume has no impact on the  enrichment factor $\varepsilon$.

This qualitative switch in the mechanism for aggregate enrichment 
raises the question which systems favor large or small compartment volumes in order to maximize aggregate enrichment $\varepsilon(\bar{\phi})$. 
Figure~\ref{fig_4_phase_diagram_large_small_compartments} depicts the regimes in terms of the reaction orders characterizing primary and secondary nucleation, $n_1$ and $n_2$, for which the maximal aggregate enrichment corresponds to smaller and larger compartment volumes. This prediction can be related to specific aggregating systems for which the  values of the reaction orders $n_1$ and $n_2$ have been  experimentally determined  (References see  caption of Fig.~\ref{fig_4_phase_diagram_large_small_compartments}). Using these values for the reaction orders, our model predicts that largest enrichment is obtained for large compartments in systems of aggregating tau and yeast prion Ure2p. These two examples  belong primarily to the class of systems where the mechanism responsible for the formation of new aggregates in the late stage is fragmentation which has a zero secondary reaction order, $n_2=0$ (i.e., nucleation is monomer independent). For non-fragmenting systems with $n_2>0$, our model predicts 
different scenarios for aggregating systems: largest aggregate enrichment for large compartment volumes occurs
in the case branching systems, such as actin in the presence of the complex Arp2/3, as well as systems proliferating through monomer dependent secondary nucleation with $n_2<n_1-1$, such as the Islet Amyloid Polypeptide (IAPP).
In contrast, largest aggregate enrichment is reached for small compartments in the case of the 40- and 42-residue forms of Amyloid-$\beta$ peptide (A$\beta$40 and A$\beta$42).

 \section*{Conclusion}
 
	By combining the theories of irreversible protein aggregation kinetics and phase separation  we have shown how liquid compartments can control the spatial pattern of irreversible protein aggregation. The coupling of slow aggregation and rapid phase separation leads to a mechanism whereby even a weak partitioning  of monomers in the case of weak protein-protein interactions is amplified into a relatively large accumulation of aggregates in the compartment. The resulting degree of aggregate enrichment is determined by two physical parameters related to phase separation (monomer enrichment $\Gamma$ and partitioning degree $\xi$) and two parameters characterizing the aggregation kinetics (reaction orders $n_1$ and $n_2$ for primary and secondary nucleation).
Since the kinetic parameters are fixed by the nature of the nucleation reactions, the phase separation parameters that are governed by the molecular interactions between the constituents  thus represent ideal control variables to regulate the degree of aggregate enrichment.

Our model may already provide a framework to explain the phenomena of aggregate partitioning inside living cells. 
An example of such phenomena could be the partitioning of pericentriolar material into centrosomes~\cite{zwicker2014centrosomes} and the spatial organization of aggregates inside stress granules~\cite{molliex2015phase,mateju2017aberrant}. The propensity of aggregates to solidify the compartment as reported in Ref.~\cite{mateju2017aberrant} could be accounted for in our model through a gel-sol transition~\cite{stockmayer1943theory,harmon2017intrinsically}. Including the solidification induced by aggregates could lead to additional volume changes of the compartment which in turn may affect the aggregation kinetics.

Our quantitative predictions of strong aggregate enrichment inside a liquid compartment (Fig.~\ref{fig_3_monomer_volume_enrichment} (b-e)) and how compartment volume affects this enrichment for different aggregating systems (Fig.~\ref{fig_4_phase_diagram_large_small_compartments}) suggest direct experimental tests for instance
using an Amyloid-$\beta$ aggregation assay~\cite{knowles2011observation} in systems where one can enhance the thermodynamic stability of droplets using surfactant-stabilized emulsions. The condition of negligible aggregate diffusion can be realized using a gel matrix with pore sizes well above the monomer size but much lower than the expected average fibril length. The monomer enrichment $\Gamma$ could be varied \emph{in vitro} by changing parameters such as salt composition, pH
or temperature, however, verifying that these changes only weakly affect the phase separation itself.

Furthermore, the enrichment of toxic aggregates inside liquid compartments  suggests new directions for drug design against aberrant protein aggregation. Our results suggest to design drugs not only with respect to their ability to interfere with the aggregation kinetics~\cite{arosio2014chemical} but also with respect to their  partitioning properties into the liquid compartments. This strategy is reminiscent of quantifying the potency of low molecular weighted anaesthetics  by the Meyer-Overton correlation~\cite{meyer1937contributions,franks1978general}.

The reported feedback mechanism of aggregate growth mediated by liquid compartments may represent a  general principle to spatially confine and speed up other irreversible chemical processes. Examples may include precipitation of proteins or polymerization kinetics of actin and microtubules (see also Fig.~\ref{fig_4_phase_diagram_large_small_compartments}). Indeed, a speed up of the chemical reactions could be expected due to the increased concentration of educts inside the liquid compartments.   Thus liquid compartments are ideal biomolecular microreactors that enrich the amount of products  by dynamically exchanging reactants with their surroundings.

 In summary, our model suggests that liquid compartments have a propensity to enrich aggregates and thus could play a role in controlling them spatially.  In our model we have considered the case   when monomers and aggregates do not affect phase separation and when phase separation is driven by the competition between the entropic tendency to mix and interactions favoring demixing.  Future work could be devoted to extending our model by entropically driven phase separation, relevant for the assembly of coacervates~\cite{overbeek1957phase}  or mixtures with depletion interactions, or to including in our model a coupling between aggregates and the liquid compartment. Moreover, our model is restricted to time scales when aggregates hardly diffuse.  Diffusion of aggregates may diminish the observed  strong aggregate enrichment  suggesting new avenues for further studies. A lowered enrichment could also be caused by the coarsening dynamics of many droplets~\cite{ostwald1897studien,Lifshitz_Slyozov_61,Bray_Review_1994}. While coarsening via coalescence would not affect our results at all because aggregates remain confined inside the droplets, we leave it to future work to understand to which extent dissolving droplets undergoing Ostwald ripening would diminish the degree of aggregate enrichment.\\

\textbf{Acknowledgements}\\
{CAW thanks the German Research Foundation (DFG) for financial support and TCTM acknowledges the support from the Swiss National Science foundation.}

\cleardoublepage
\section*{Supplemental Information}

 \section{Partition coefficient for dilute monomers at equilibrium}\label{app_partition_coefficient}

We consider phase separation of an incompressible, ternary mixture composed of monomers, component $A$ and $B$ (we neglect the interactions between the aggregates and phase separation for simplicity) 
 described by the following Flory-Huggins free energy density~\cite{flory1942thermodynamics,huggins42}
\begin{align}\nonumber
	f  &=k_B T \bigg[  \frac{\phi_A}{\nu_A}  \ln (\phi_A) +  \frac{\phi_B}{\nu_B}  \ln (\phi_B) +  c_{\rm m} \ln (\nu_{\rm m} c_{\rm m}) \\
	&\quad  +  \Lambda  \, \phi_A \phi_B 
	+ c_{\rm m} \nu_{\rm m} \left( \Lambda_{mA} \phi_A + \Lambda_{mB} \phi_B \right) \bigg] \, ,
	  \label{eq:free_energy_ternary_functional_theory_hom} 
\end{align}
where the logarithmic terms correspond to the mixing entropy. 
The interactions between $A$ and $B$ are described by the parameter $\Lambda$ while the interactions with the monomers are characterized by $\Lambda_{mi}$, $i=A,B$.
These interaction parameters have the unit $[1/\text{volume}]$. 
Here we define a dimensionless interaction parameters by writing $\Lambda =\chi/\nu$ and $\Lambda_{mi} =\chi_{mi}/\nu$ with $\nu=\nu_A$. Moreover, for simplicity, we consider equal molecular volumes of $A$ and $B$, $\nu=\nu_B$.
Thus we arrive at
\begin{align}\nonumber
	f  &=\frac{k_B T}{\nu} \bigg[  \phi_A \ln (\phi_A) +  
	 \phi_B \ln (\phi_B) + \nu c_{\rm m} \ln (\nu_{\rm m} c_{\rm m}) \\
	&\quad  +  \chi  \, \phi_A \phi_B 
	+ c_{\rm m} \nu_{\rm m} \left( \chi_{mA} \phi_A + \chi_{mB} \phi_B \right) \bigg] \, .
	  \label{eq:free_energy_ternary_functional_theory_hom2} 
\end{align}
With $\phi_B=1-\phi_A-\nu_{\rm m} c_{\rm m}$  and monomers being dilute, $\nu_{\rm m} c_{\rm m} \ll 1$, 
we can expand $f$ in  $\nu_{\rm m} c_{\rm m}$ up to the first order:
\begin{align}\nonumber
	f(\phi, c_{\rm m})  &\simeq \frac{k_B T}{\nu} \bigg\{  \phi  \ln (\phi) +   (1-\phi  ) \ln (1-\phi) +   \chi  \, \phi (1-\phi) \\
	\nonumber
	&\quad +  \nu_{\rm m} c_{\rm m} \bigg( (\nu/\nu_{\rm m}) \ln (\nu_{\rm m} c_{\rm m}) - \ln (1-\phi) \\
	&\quad   
	+   \left(\chi_{mA}-\chi_{mB} -\chi\right)  \phi + \chi_{mB} - 1\bigg) \bigg\} \, ,
	  \label{eq:free_energy_ternary_functional_theory_hom} 
\end{align}
where we defined $\phi_A\equiv \phi$ and neglected all term of order $\mathcal{O}\left[ \left( \nu_{\rm m} c_{\rm m} \right)^2\right]$.
The corresponding chemical potentials read
\begin{subequations} \label{eq:chem_pots}
\begin{align} \label{eq:chem_pot_A}
	\tilde \mu(\phi, c_{\rm m})  &= \nu \frac{\partial f}{\partial \phi} \\
	\nonumber
	&= k_B T \bigg[ 
	 \ln { \phi}  -  \ln { (1- \phi)}  + \chi (1-2 \phi)  \\
	\nonumber
	& \quad +  c_{\rm m} \nu_{\rm m} \left(\chi_{mA}-\chi - \chi_{mB} +  \frac{1}{1-\phi}\right) 
	\bigg] \, ,
	  \\
	   \label{eq:chem_pot_m}
	  	   \mu_{\rm m}(\phi, c_{\rm m})  &=  \frac{\partial f}{\partial c_{\rm m}} \\
		   \nonumber
		   &= k_B T \bigg\{ 
	   \ln { (\nu_{\rm m} c_{\rm m}) } + 1 + \frac{\nu_{\rm m}}{\nu} \ln (1-\phi)   \\
	   \nonumber
	   & \quad  +
	   \frac{\nu_{\rm m}}{\nu} \bigg[ \left(\chi_{mA}-\chi\right)  \phi + \chi_{mB} (1-\phi) - 1\bigg]
 \bigg\} \, .
\end{align}
\end{subequations}
At equilibrium  the chemical potentials of each component  inside (I) and outside (II) the compartment are balanced leading to relationships between the concentration values inside and outside.
Specifically, if phase separation equilibrium is reached the following conditions are fulfilled 
\begin{subequations}
\begin{align}
\label{eq_equil_A}
	\tilde{\mu}^\text{I}(\phi^\text{I}, c_{\rm m}^\text{I}) &=\tilde{\mu}^\text{II}(\phi^\text{II}, c_{\rm m}^\text{II}) \, ,\\
\label{eq_equil_m}
	\mu_{\rm m}^\text{I}(\phi^\text{I}, c_{\rm m}^\text{I})&=\mu_{\rm m}^\text{II}(\phi^\text{II}, c_{\rm m}^\text{II}) \, .
\end{align}
\end{subequations}
The relations above allow to calculate the equilibrium concentration in each phase, for component $A$, $\phi^\text{I}$ and $\phi^\text{II}$, and the monomers, $c_{\rm m}^\text{I}$ and $c_{\rm m}^\text{II}$.
An analytic result of the equilibrium concentrations is very difficult to obtain.
However, we can focus on the leading contributions for the balance of the chemical potentials inside and outside taking advantage that monomers are dilute and thereby obtain an approximation for the equilibrium values inside and outside. 

Considering that the dimensionless interaction parameters $\chi$ are all of  $\mathcal{O}(1)$, the impact of the dilute monomers on the phase equilibrium between $A$ and $B$ is negligible and we can approximate 
\begin{align} \label{eq:chem_pot_A_approx}
	\tilde \mu(\phi, c_{\rm m})  & \simeq k_B T \bigg[ 
	 \ln {\left( \frac{\phi}{1-\phi} \right)} + \chi (1-2 \phi) 	\bigg] \, .
\end{align}
 The resulting chemical potential Eq.~\eqref{eq:chem_pot_A_approx} simply corresponds to the chemical 
potential of a binary, incompressible Flory-Huggins mixture.
From the equilibrium condition Eq.~\eqref{eq_equil_A}, we can calculate the binodal line described by the condition
\begin{equation}
	\chi \simeq \frac{ \ln {\left( {\phi/(1-\phi)} \right)}}{2 \phi -1} \, ,
\end{equation}
 which solely depends on the interaction parameters between $A$ and $B$, $\chi$.
By means of equilibrium condition Eq.~\eqref{eq_equil_m}, we can calculate the impact of the phase separated compartment on the monomer distribution leading to 
the monomer enrichment
 \begin{equation}\label{relation_Gamma}
\frac{c_{\rm m}^\textrm{I}}{c_{\rm m}^\textrm{II}}
\simeq  \exp\left[\frac{\nu_{\rm m}}{\nu}\Delta \chi  \left( \phi^\textrm{I} - \phi^\textrm{II}\right) \right] =: \Gamma
 \, ,
\end{equation}
where the relative interaction strength reads
 $\Delta \chi=\left(\chi_{B,\textrm{m}}  -\chi_{A,\textrm{m}} \right)$.
 Thus, there is enrichment of monomers in the condensed phase ($\Gamma>1$) if monomers favor the presence of $A$ relative to $B$, i.e., $\chi_{A,\textrm{m}} < \chi_{B,\textrm{m}}$.

 \subsection*{Inter-compartment flux of monomers close to
equilibrium}\label{app_flux}


 This flux can be calculated for a maintained concentration difference $\phi^\text{I} - \phi^\text{II}$ using the chemical potential for the monomers $\mu_{\rm m}$ (Eq.~\eqref{eq:chem_pot_m}).
 To this end, let us assume that the compartment is spherical with a radius $R$ and corresponding volume $V^\text{I}=(4\pi/3)R^3$.
 Perturbing the concentrations in both phases may lead to an unbalance of the chemical potential and thus a flux between the phases.
The total change of monomer due to a net flux through the interface between the compartments reads:
  \begin{align}
 	J =  R^2 \lim_{\Delta x \to 0}\int \text{d}\varphi \,  \text{d}\theta \,  \vect{e}_r \cdot \vect{j}\big|_{R+\Delta x/2}^{R-\Delta x/2}   \, ,
 \end{align}
 where the unbalance of the chemical potential occurs through the interface of width $\Delta x$ and $\vect{e}_r$ is the radial unit vector pointing normal to the spherical interface.
The local flux  $\vect{j}$ can be approximated as  $\vect{j}= - \xi  \nabla \mu_\text{m} \simeq - \xi  \vect{e}_r \partial_r \mu_\text{m} \simeq -  \vect{e}_r \xi (\mu_\text{m}^\text{I}-\mu_\text{m}^\text{II}) /\Delta x$, where $\xi$ denotes the mobility coefficient. 
This approximation neglects spatial inhomogeneities in chemical potentials within the phases which is valid in the limit of fast diffusion on the length scale of the system of volume $V$.
 Using the chemical potential of the monomers, Eq.~\eqref{eq:chem_pot_m},
 we can approximate the gradient of the chemical potential as
    \begin{align}
 	\frac{\mu_\text{m}^\text{I}-\mu_\text{m}^\text{II}}{\Delta x } 
	& =  \frac{k_BT}{\Delta x} \left[  \ln{\left( \mamI \right)} - \ln{\left( \Gamma \, \mamII  \right)} 
	 \right] \\
	   & \simeq  \frac{k_BT}{\Delta x} \frac{ \delta \mamI - \Gamma \,  \delta \mamII }{\mamIeq}
	   \, ,
 \end{align}
 where we expanded $\mamI=\mamIeq + \delta \mamI$ and $\mamII= \mamIIeq + \delta \mamII$  up to linear order around the equilibrium concentrations $\mamIeq$ and $\mamIIeq=\mamIeq/\Gamma$, respectively. 
Thus the change in concentration due to the exchange of material through the interface reads
   \begin{align}
 	\frac{J}{V^{(\alpha)}(R)} & \simeq - \frac{4 \pi R^2}{V^{(\alpha)}(R) } \frac{ D_{\rm m} }{\Delta x }  \, \frac{   \mamI - \Gamma  \mamII }{\nu_\text{m}\, \mamIeq}  
	\\
	& = - k^{(\alpha)}(R) \,  \left(  \mamI - \Gamma  \mamII \right) \, ,
 \end{align}
  where the diffusion constant  $D_{\rm m} = k_bT \nu_\text{m}\xi$. To ease notation we omitted the ``$\delta$'' to indicate linear deviations from  equilibrium. 
  Moreover,  the rate to relax back to monomer partitioning at equilibrium is 
  \begin{equation}\label{eq_rate_mon_equil}
  k^\alpha(R)=\frac{4 \pi R^2 D_{\rm m}}{\Delta x  V^{(\alpha)}(R)\nu_\text{m}\mamIeq} \, .
  \end{equation}
This rate depends on parameters such as the monomer diffusion constant  and the size of the compartment $V^\text{I}$ (Eq.~\eqref{eq_rate_mon_equil}).
In particular, for large compartment I, $V^\text{I}\approx V$, the fraction of rates, $k^\text{I}/k^\text{II}$, decreases toward zero indicating that the relaxation in compartment I is slowed down relative to compartment II. Conversely, in the case of small compartments, $V^\text{I}\ll V$, relaxation of compartment I is fast compared to compartment II. 
Most importantly, the size of the compartment does not affect the equilibrium concentration.
  For simplicity, we neglect the impact of surface tension which leads 
   to a weak increase of the equilibrium volume fractions.

\section{Analytical solution to aggregation kinetics with liquid compartments}
\label{analytics_agg_meets_PS}

In this appendix, we discuss the detail associated with the derivation of analytical solutions to the aggregation kinetics in the presence of liquid compartments, Eq.(3), main text.


\subsection{Boundary layer dynamics}

Due to the separation of timescales between monomer equilibration between the two compartments and protein aggregation, the system described by Eqs.~(3) (main text) will develop initially through a rapid phase of equilibration (boundary layer), before any aggregation occurs in either compartment. During this phase, the initial values of the monomer concentration in each compartment, $M_{\rm{m}}^\text{I}(0)$ and $M_{\rm{m}}^\text{II}(0)$, relax quickly to equilibrium such that the condition $M_{\rm{m}}^\text{I}(t)=\Gamma M_{\rm{m}}^\text{II}(t)$ is satisfied before aggregation is initiated. This early equilibration kinetics is described by setting the aggregation terms in Eqs.~(3) (main text) to zero, yielding the following equations:
\begin{subequations}\label{1234}
\begin{align}
\frac{\text{d}M_{\rm a}^\text{I} (t)}{\text{d}t} &= - k^\text{I} \,\big[ M_{\rm{m}}^\text{I}(t) -\Gamma M_{\rm{m}}^\text{II}(t)\big] 
 \, ,\\
\frac{\text{d}M_{\rm a}^\text{II} (t)}{\text{d}t} &=  k^\text{II} \,\big[ M_{\rm{m}}^\text{I}(t) -\Gamma M_{\rm{m}}^\text{II}(t)\big] 
\, .
 \end{align} 
 \end{subequations}
 The solution to Eqs.~\eqref{1234} is:
 \begin{subequations}\label{abcd}
\begin{align}
M_{\rm m}^\text{I} &= \Gamma B + \frac{k^\text{I} \,\big[ M_{\rm{m}}^\text{I}(0) -\Gamma M_{\rm{m}}^\text{II}(0)\big] }{A}  e^{-A t}
 \, ,\\
M_{\rm m}^\text{II} &=   B+ \frac{k^\text{II} \,\big[ M_{\rm{m}}^\text{I}(0) -\Gamma M_{\rm{m}}^\text{II}(0)\big] }{A}  e^{-A t}
\, ,
 \end{align} 
 \end{subequations}
 where $A = k^\text{I}+\Gamma k^\text{II}$, and $B = \big[ k^\text{II} M_{\rm{m}}^\text{I}(0) +k^\text{I} M_{\rm{m}}^\text{II}(0)\big]/A$. Note that the kinetics described by Eq.~\eqref{abcd} `pushes' the system towards the slow manifold, which is described by $M_{\rm{m}}^\text{I}(t)=\Gamma M_{\rm{m}}^\text{II}(t)$. Hence, when $M_{\rm{m}}^\text{I}(0) =\Gamma M_{\rm{m}}^\text{II}(0)$, there is no initial phase of `correction' of the initial conditions. At the end of this initial boundary layer phase, the monomer concentrations in the two compartments are given by: 
 \begin{subequations}
\begin{align}
M_{\rm m}^\text{I} &= \Gamma B
 \, ,\\
M_{\rm m}^\text{II} &=   B
\, .
 \end{align} 
 \end{subequations}
Since we are not very much interested in this initial phase of redistribution of the initial conditions, in the following we shall assume for simplicity that the initial monomer concentrations in compartments I and II satisfy the relationship $M_{\rm{m}}^\text{I}(0) =\Gamma M_{\rm{m}}^\text{II}(0)$. This assumption does not affect the generality of our results. In fact, if this condition was not satisfied initially, then, according to Eq.~\eqref{abcd}, rapid equilibration between the two compartments would correct these initial conditions, eventually leading to a 'corrected' set of initial conditions that lie in the slow manifold.
 
\subsection*{Solving aggregation kinetics in the slow manifold}

\begin{figure}
	\begin{tabular}{cc}
		\includegraphics[width=0.25\textwidth]{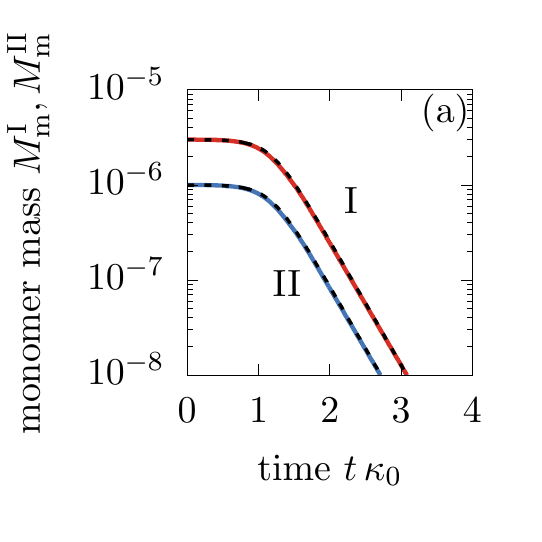} &
		\includegraphics[width=0.25\textwidth]{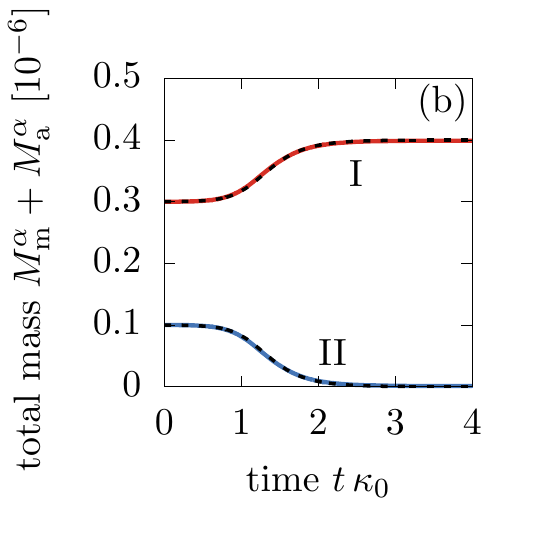}
\end{tabular}
	\caption{Comparison between the analytical solutions Eqs.~\eqref{sol_mon}, \eqref{sol_agg}, \eqref{sol_monII} (dashed lines) and the numerical solution to Eqs.~(3) in the main text (solid lines). The parameters are: $k_+=10^{6}$ M$^{-1}$s$^{-1}$, $k_1=10^{-4}$ M$^{-1}$s$^{-1}$, $k_2=10^{4}$ M$^{-2}$s$^{-1}$, $M_{\rm{m}}^{\rm{tot}}=1$ $\mu$M, $n_1=n_2=2$, $\Gamma=3$, $\xi=1$ and $k^\alpha/\kappa_0=100$ for $\alpha=\text{I,II}$. 
\label{SIfig1}
}
\end{figure}

After an initial, rapid phase of monomer redistribution through the two compartments, the system enters a slower phase of dynamics, where, at leading order, the system stays on the slow manifold $\mamI= \Gamma \mamII$ at all times. For simplicity, let us assume that the initial concentrations of monomers in the two compartments obey the relationship $M_{\rm{m}}^\text{I}(0) =\Gamma M_{\rm{m}}^\text{II}(0)$ (otherwise there will be a fast equilibration of the initial conditions such that this relationship is satisfied). To describe the aggregation process in the slow manifold, we write $\mamI- \Gamma \mamII\simeq 0$ for all times in Eqs.~(3) (main text) and find: 
 \begin{subequations}\label{1133slow}
\begin{align}
\label{1133aslow}
\frac{\text{d}\mamI (t)}{\text{d}t} &= -2k_+\, \mamI(t) \, \cpI(t) = - \frac{\text{d}M_{\rm a}^\text{I}(t)}{\text{d}t} \, ,\\
\frac{\text{d}\mamII (t)}{\text{d}t} &= -2k_+\, \mamII(t) \, \cpII(t) =- \frac{\text{d}M_{\rm a}^\text{II}(t)}{\text{d}t}  \, ,\\ 
\label{1133bslow}
\frac{\text{d}c_{\rm a}^\text{I}(t)}{\text{d}t} & = k_1 \, M_{\rm m}^\text{I}(t)^{n_1}+k_2 \, M_{\rm m}^\text{I}(t)^{n_2} \, M_{\rm a}^\text{I}(t) \, ,\\ 
\frac{\text{d}c_{\rm a}^\text{II}(t)}{\text{d}t} & = k_1 \, M_{\rm m}^\text{II}(t)^{n_1}+k_2 \, M_{\rm m}^\text{II}(t)^{n_2} \, M_{\rm a}^\text{II}(t) \, .
 \end{align} 
 \end{subequations}
It is useful to introduce the total monomer concentration in the system as:
\begin{equation}
M_{\rm{m}}(t)  =  \frac{ V^\text{I} \mamI (t) + V^\text{II} \mamII (t)}{V}.
\end{equation} 
Note that this concentration may vary in time as aggregates are nucleated and grow, however, the total mass concentration 
\begin{equation}
M_{\rm{m}}^{\rm{tot}}  = M_{\rm{m}}(t)  + \frac{ V^\text{I} M_{\rm a}^\text{I} (t) + V^\text{II} M_{\rm a}^\text{II}(t)}{V}\, 
\end{equation} 
is conserved at all times.
Using the condition $\mamI= \Gamma \mamII$, we can write the following relationships linking the concentrations of monomers in compartments I and II to the total concentration of monomers in the system:
  \begin{subequations}\label{MIandMtot}
  \begin{align}
 \mamI (t) & = \xi \,  \Gamma  M_{\rm{m}}(t) \\
 \mamII (t) & = \xi \, M_{\rm{m}}(t) \, ,
 \end{align}
 \end{subequations}
 where we abbreviated the partitioning degree
 \begin{equation}\label{eq_g}
 	\xi=\frac{V}{(\Gamma -1)V^\text{I}+V} .
\end{equation}
Thus, using Eqs.~\eqref{MIandMtot}, we can re-write Eqs.~\eqref{1133slow} as:
\begin{subequations}\label{11133fast_intercompartment_kinetics}
\begin{align}
\label{11133a}
\frac{\text{d}M_{\rm a}^\text{I}(t)}{\text{d}t} &= 2k_+( \xi \,  \Gamma) M_{\rm{m}}(t)  \cpI(t) 
 \, ,\\
\frac{\text{d}M_{\rm a}^\text{II}(t)}{\text{d}t} &= 2k_+ \xi  M_{\rm{m}}(t) \cpII(t)
\, ,\\
 \frac{\text{d}\cpI(t)}{\text{d}t} & =k_1 (\xi \,  \Gamma)^{n_1} M_{\rm{m}}(t)^{n_1}+k_2 (\xi \,  \Gamma)^{n_2}  M_{\rm{m}}(t)^{n_2}  M_{\rm a}^\text{I}(t)  \, ,\\
\label{11133b}
\frac{\text{d}\cpII(t)}{\text{d}t} & =k_1 \xi^{n_1} M_{\rm{m}}(t)^{n_1}+k_2 \xi^{n_2} M_{\rm{m}}(t)^{n_2} M_{\rm a}^\text{II}(t)  \, .
 \end{align} 
 \end{subequations}

\subsubsection{Early-time dynamics for aggregate number and mass concentrations in compartments I and II}

Before discussing the full time course of aggregation, it is useful to consider the early-time kinetics of the system, which emerges when the monomers in the system have not been depleted significantly \cite{michaels2018chemical}. This limit is obtained by assuming in \eqref{11133fast_intercompartment_kinetics} that the total monomer concentration is constant in time, i.e., $M_{\rm{m}}(t)\approx M_{\rm{m}}^{\rm{tot}}$ \cite{michaels2018chemical}. This assumption transforms the kinetic equations \eqref{11133fast_intercompartment_kinetics} into the following simpler set of linear differential equations:
\begin{subequations}\label{11133lin}
\begin{align}
\label{11133lina}
\frac{\text{d}M_{\rm a}^\text{I}(t)}{\text{d}t} &= \mu_{\rm{I}} \, \cpI(t)  \, ,\\
\frac{\text{d}M_{\rm a}^\text{II}(t)}{\text{d}t} &= \mu_{\rm{II}} \, \cpII(t)  \, ,\\
 \frac{\text{d}\cpI(t)}{\text{d}t} & =\nu_{\rm{I}}+\beta_{\rm{I}} \, M_{\rm a}^\text{I}(t)  \, ,\\
\label{11133linb}
\frac{\text{d}\cpII(t)}{\text{d}t} & =\nu_{\rm{II}}+\beta_{\rm{II}} \, M_{\rm a}^\text{II}(t)  \, ,
 \end{align} 
 \end{subequations}
 where we have introduced the parameters:
\begin{align}
 \mu_{\rm{I}} &= \mu_0 \,  \xi \,  \Gamma \, ,   &\nu_{\rm{I}} = \nu_0 (\xi \,  \Gamma)^{n_1}  \, , &  \quad \beta_{\rm{I}} =\beta_0 ( \xi \,  \Gamma)^{n_2}  \, ,
 \end{align} 
 and
 \begin{align}
  \mu_{\rm{II}} &=  \mu_0 \, \xi \, ,   &\nu_{\rm{II}} = \nu_0 \,  \xi^{n_1}  \, , &  \quad \beta_{\rm{II}} = \beta_0 \, \xi^{n_2}  \,  ,
 \end{align} 
 with
 \begin{subequations}
\begin{align} \nu_0 & =k_1 (\mamt)^{n_1}\, ,\\
  \rb_0 & =k_2 (\mamt)^{n_2}\, ,\\
   \mu_0 & =2k_+\mamt. \end{align} 
 \end{subequations}
 The solution to Eqs.~\eqref{11133lin} subject to the condition that no aggregates are present initially reads
 \begin{subequations}\label{agg}
\begin{align}
 \frac{M_{\rm a}^\text{I}(t)}{M_{\rm m}^\text{I}(0)} & = \frac{\lambda_\text{I}^2[\cosh(\kappa_\text{I}t)-1]}{\kappa_\text{I}^2},\\
\frac{ M_{\rm a}^\text{II}(t)}{M_{\rm m}^\text{II}(0)} & = \frac{\lambda_\text{II}^2[\cosh(\kappa_\text{II}t)-1]}{\kappa_\text{II}^2} \, ,
 \end{align} 
 \end{subequations}
 for the aggregate mass concentrations in compartments I and II, and
 \begin{subequations}\label{56}
\begin{align}
  c_{\rm a}^\text{I}(t) & = \frac{\nu_\text{I}\sinh(\kappa_\text{I}t)}{\kappa_\text{I}}\\
c_{\rm a}^\text{II}(t) & = \frac{\nu_\text{II}\sinh(\kappa_\text{II}t)}{\kappa_\text{II}}
 \end{align} 
 \end{subequations}
 for the aggregate number concentrations in compartments I and II. Here, we have introduced the kinetic coefficients
\begin{equation}
\lambda_{\rm{I}} =\lambda_0 \,  (\xi  \Gamma)^{\frac{n_1}{2}}  \, , \quad \lambda_\text{II} =\lambda_0 \,  \xi^{\frac{n_1}{2}}  \, ,
\end{equation}
and
\begin{equation}
\kappa_{\rm{I}} =\kappa_0 \,  (\xi  \Gamma)^{\frac{n_2+1}{2}}  \, , \quad \kappa_\text{II} =\kappa_0 \,  \xi^{\frac{n_2+1}{2}}  \, ,
\end{equation}
with 
\begin{subequations}
\begin{align}
\lambda_0& =\sqrt{2k_+k_1 (M_{\rm{m}}^{\rm{tot}})^{n_1}} \, ,\\
\kappa_0 & =\sqrt{2k_+k_2 (M_{\rm{m}}^{\rm{tot}})^{n_2+1}} \, ,
\end{align}
 \end{subequations}
 being the effective rates characterizing the proliferation of aggregates due to primary and secondary nucleation, respectively \cite{michaels2018chemical}.
According to Eqs.~\eqref{agg} and \eqref{56}, the aggregate number and mass concentrations in both compartments grow exponentially with time. The effective growth rates $\kappa_{\rm{I}}$ and $\kappa_{\rm{II}}$ are different for each compartment and depend on $\Gamma$ and $\xi$. Since $\Gamma\gg 1$, aggregate growth in the early times is much faster in compartment I compared to compartment II. In particular, the ratio of the growth rates in the two compartments is independent of $\xi$ and is given by: 
\begin{equation}\label{kioverkii}
	\frac{\kappa_{\rm{I}}}{\kappa_\text{II}} = \Gamma^{\frac{n_2+1}{2}} \, .
\end{equation}
Also primary nucleation is enhanced inside compartment I relative to compartment II:
\begin{equation}\label{kioverkii2}
	\frac{\nu_{\rm{I}}}{\nu_{\rm{II}} }= \Gamma^{n_1} \, .
\end{equation}

 \subsubsection{Analytical solution for full time course of monomer concentrations in compartments I and II}
 
We now construct analytical solutions for the monomer and aggregate mass concentrations that are valid for the entire duration of the aggregation reaction. In the previous section, we have seen that for $\Gamma\gg 1$ two timescales, $1/\kappa_{\rm{I}}$ and $1/\kappa_\text{II}$, characterize the early-time aggregation in the two compartments. Since the growth rate in compartment I, $\kappa_{\rm{I}}$, is much larger than that in compartment II, $\kappa_{\rm{II}}$, monomers in compartment I will be consumed by aggregation much faster than those in compartment II. However, the relationship $M_{\rm m}^\text{I}(t)=\Gamma M_{\rm m}^\text{II}(t)$ must hold at all times. Thus, to compensate the fast aggregation in compartment I, there will be a flux of monomers from compartment II to compartment I. Eventually, the vast majority of monomers will end up as part of aggregates in compartment I and the parameter $\kappa_{\rm{I}}$ will naturally control the depletion of monomers in both compartments. We can make this argument more quantitative by using Eqs.~\eqref{agg} as follows. Monomers in compartment I are consumed over a timescale of the order $1/\kappa_{\rm{I}}$. The amount of aggregate mass that will be formed in compartment II during this time period will be of the order
\begin{equation}
M_{\rm a}^\text{II}  \simeq M_{\rm m}^\text{II}(0) \frac{\lambda_\text{II}^2[\cosh(\kappa_\text{II}/\kappa_{\rm{I}})-1]}{\kappa_\text{II}^2}.
\end{equation} Since $\kappa_{\rm{II}}/\kappa_{\rm{I}}\ll 1$, we can expand the cosh function as a Taylor series, $\cosh x = 1+x^2/2+\mathcal{O}(x^5)$. At leading order, we find:
\begin{equation}
M_{\rm a}^\text{II}  \simeq M_{\rm m}^\text{II}(0) \frac{\lambda_\text{II}^2}{2\kappa_\text{I}^2}.
\end{equation}
Thus, the ratio between the mass of aggregates formed in compartments I and II over a timescale $1/\kappa_{\rm{I}}$ is
\begin{equation}\label{sca}
\frac{M_{\rm a}^\text{I} }{M_{\rm a}^\text{II} } \simeq \frac{M_{\rm a}^\text{I}(0) }{M_{\rm a}^\text{II}(0) }\frac{\lambda_\text{I}^2}{\lambda_\text{II}^2}\simeq \Gamma^{\frac{n_1}{2}+1}.
\end{equation}
Since $\Gamma \gg 1$, the aggregate mass in compartment I will be much larger than that in compartment II. 
We can thus neglect at leading order the contribution from $M_{\rm{a}}^{\rm{II}}(t)$ to the conservation of total mass relationship. Doing so, and using Eqs.~\eqref{MIandMtot}, we can write the conservation of mass relationship as follows
\begin{equation}\label{sdf}
M_{\rm{a}}^{\rm{I}}(t) = \big[M_{\rm m}^\text{I}(0)-M_{\rm m}^\text{I}(t)\big] \frac{\Gamma+1}{\Gamma}.
\end{equation}
Using Eq.~\eqref{sdf}, we can reduce the kinetic equations \eqref{11133fast_intercompartment_kinetics} to a system of two coupled different equations:
  \begin{subequations}\label{1133slow2}
\begin{align}
\label{1133aslow2}
\frac{\text{d}\mamI (t)}{\text{d}t} &= -2\tilde{k}_+\, \mamI(t) \, \cpI(t)  \, ,\\
\label{1133bslow2}
\frac{\text{d}c_{\rm a}^\text{I}(t)}{\text{d}t} & = k_1 \, M_{\rm m}^\text{I}(t)^{n_1}+\tilde{k}_2 \, M_{\rm m}^\text{I}(t)^{n_2} \, \big[M_{\rm m}^\text{I}(0)-M_{\rm m}^\text{I}(t)\big]  \, ,
 \end{align} 
 \end{subequations}
 where $\tilde{k}_+ = k_+ \Gamma/(\Gamma+1)$ and $\tilde{k}_2 = k_2 (\Gamma+1)/\Gamma$.
Conveniently, Eqs.~\eqref{1133slow2} are exactly the fundamental kinetic equations describing the dynamics of protein aggregation in a pure system, i.e., without compartment, but with effective rate parameters that depend on the degree of phase separation \cite{michaels2018chemical,michaels2016hamiltonian}. Thus, we can adapting the results in \cite{michaels2016hamiltonian} to Eq.~\eqref{1133slow2}, we find the following solution for the time varying monomer concentration in compartment I:
\begin{align}\label{sol_mon}
\frac{\mamI(t)}{M_{\rm m}^\text{I}(0)} &= \left[1+\frac{\lambda_{\rm{I}}^2}{2\kappa_{\rm{I}}^2\theta}\left(\frac{\Gamma}{\Gamma+1}\right)e^{\kappa_{\rm{I}} t}\right]^{-\theta} \, ,
 \end{align} 
  where $\theta=\sqrt{2/[n_2(n_2+1)]}$. Using Eq.~\eqref{sdf}, we then obtain an expression for the aggregate mass concentration:
  \begin{align}\label{sol_agg}
\frac{M_{\rm a}^\text{I}(t)}{M_{\rm m}^\text{I}(0)} &=\frac{\Gamma+1}{\Gamma}\left(1- \left[1+\frac{\lambda_{\rm{I}}^2}{2\kappa_{\rm{I}}^2\theta}\left(\frac{\Gamma}{\Gamma+1}\right)e^{\kappa_{\rm{I}} t}\right]^{-\theta}\right) \, .
 \end{align} 
 Finally, the time course of the monomer concentration in compartment II is obtained using the relationship $M_{\rm m}^\text{I}(t)=\Gamma M_{\rm m}^\text{II}(t)$. This yields:
  \begin{equation}\label{sol_monII}
 \frac{\mamII(t)}{M_{\rm m}^\text{II}(0)} = \left[1+\frac{\lambda_{\rm{I}}^2}{2\kappa_{\rm{I}}^2\theta}\left(\frac{\Gamma}{\Gamma+1}\right)e^{\kappa_{\rm{I}} t}\right]^{-\theta} \, . 
 \end{equation}
 The accuracy of our analytical solutions Eqs.~\eqref{sol_mon}, \eqref{sol_agg} and \eqref{sol_monII} against numerical integration of Eqs.~(3) (main text) is shown in Fig.~\ref{SIfig1}.

\subsubsection{Scaling relationships for the aggregate number concentrations in compartments I and II}

From the knowledge of the time varying monomer concentration, Eq.~\eqref{sol_mon}, we can obtain an expression for the aggregate number concentration in compartment I using Eq.~\eqref{1133aslow2} by simple differentiation of Eq.~\eqref{sol_mon}, $\cpI(t)= -1/[2\tilde{k}_+\, \mamI(t)] \text{d}\mamI (t)/\text{d}t $. This yields the following expression:
   \begin{align}
\frac{c_{\rm a}^\text{I}(t)}{c_{\rm a}^\text{I}(\infty)} & =  \left[1+\frac{2\kappa_\text{I}^2\theta}{\lambda_{\rm{I}}^2}\left(\frac{\Gamma+1}{\Gamma}\right)e^{-\kappa_{\rm{I}} t}\right]^{-1}\, ,
 \end{align} 
where \begin{equation}\label{cinf}
 c_{\rm a}^\text{I}(\infty) = \frac{ \kappa_{\rm{I}}\theta}{2k_+}\left(\frac{\Gamma+1}{\Gamma}\right)
 \end{equation}
 is the number concentration of aggregates at steady state.
It is interesting to extract from Eq.~\eqref{cinf} the key dependence of $c_{\rm a}^\text{I}(\infty)$ on the parameters $\xi$ and $\Gamma$:
  \begin{align}
 c_{\rm a}^\text{I}(\infty) & = \frac{\kappa_0\theta }{2k_+ }\xi^{\frac{n_2+1}{2}} \Gamma^{\frac{n_2-1}{2}}(\Gamma+1) \, .
 \end{align}
 Note that the prefactor defines the homogeneous concentration in the absence of compartments,
 \begin{equation}
 	c_\textrm{a}^0 = \frac{\kappa_0\theta }{2k_+} \, .
 \end{equation}
 Thus, using $\Gamma \gg 1$ we find the following scaling relationship for the steady-state number concentration of aggregates in compartment I:
 \begin{align}\label{cI}
 c_{\rm a}^\text{I}(\infty) \simeq  c_\textrm{a}^0 \,  (\xi\Gamma)^{\frac{n_2+1}{2}} .
 \end{align}
 A similar scaling relationship can be derived also for the steady-state number concentration of aggregates in compartment II as follows. We recall that the early-time dynamics of aggregation in compartment II is characterized by a timescale $1/\kappa_{\rm{II}}$, which is much slower than the timescale of aggregation in compartment I, $1/\kappa_{\rm{I}}$. 
 %
Thus, $c_{\rm a}^\text{II}$ can be considered to be still in the exponential growing phase even when the aggregate concentration in compartment I is equilibrating. Eventually, the assembly in compartment II is arrested abruptly as soon as aggregation in compartment I is fully saturated, since no monomer is left in either compartment. Since the timescale for saturation of aggregation in compartment I is proportional to $ 1/\kappa_{\rm{I}}$ (see Eq.~\eqref{sol_mon}), we can estimate the concentration of aggregates in compartment II at the end of the reaction as:
 \begin{equation}\label{sinh}
  c_{\rm a}^\text{II}(\infty) \simeq  \frac{\nu_{\rm{II}}\sinh(\kappa_{\rm{II}}/\kappa_{\rm{I}})}{\kappa_{\rm{II}}} = \frac{\nu_{\rm{II}}}{\kappa_{\rm{II}}} \sinh\left(\Gamma^{-\frac{n_2+1}{2}}\right),
 \end{equation}
 where in the last step we used Eq.~\eqref{kioverkii}.
 Since $\Gamma\gg 1$, the argument of the $\sinh$ function is much smaller than unity. Hence, we can simplify Eq.~\eqref{sinh} by using a Taylor expansion of the $\sinh$ function to first order, $\sinh x = x+\mathcal{O}(x^3)$, yielding the following scaling relationship after extracting the $\xi$ dependence of $\nu_{\rm{II}}$ and $\kappa_{\rm{II}}$:
  \begin{equation}\label{cII}
  c_{\rm a}^\text{II}(\infty) \simeq c_\textrm{a}^0 w \,  \xi^{{n_1}-\frac{n_2+1}{2}}\Gamma^{-\frac{n_2+1}{2}} \, ,
 \end{equation}
 where $w=k_1/(k_2 \theta) \left( M_{\rm m}^{\rm tot} \right)^{n_1-n_2-1}$.
 Combining Eq.~\eqref{cI} with \eqref{cII}, we obtain one of the key results of our paper, namely the scaling behavior of aggregate enrichment between compartments I and II with $\Gamma$:
   \begin{equation}
  \frac{c_{\rm a}^\text{I}(\infty)}{c_{\rm a}^\text{II}(\infty)} \propto \xi^{n_2+1-n_1}\Gamma^{n_2+1}\, ,
 \end{equation}
where the impact of compartment volume on the relative degree of monomer characterized by $\xi\left( \bar \phi \right)$
is given in Eq.~\eqref{eq_g}.


%

\end{document}